\documentclass[nofootinbib]{revtex4}
\usepackage{graphicx}
\usepackage{latexsym}
\def\be{\begin{equation}}
\def\ee{\end{equation}}
\def\bea{\begin{eqnarray}}
\def\eea{\end{eqnarray}}

\begin{document}

\title{Mass Operator and Gauge Field Theory with Five-variable Field Functions}
\author{ChiYi Chen${}^{a}$}
\email{iamchiyi@gmail.com}

 \affiliation{${}^a$HangZhou Normal University, Hangzhou 310036, China}

\begin{abstract}
To investigate the mass generating problem without Higgs mechanism we present a model in which a new scalar gauge coupling is naturally introduced. Because of the existence of production and annihilation for particles in quantum field theory, we extend the number of independent variables from conventional four space-time dimensions to five ones in order to describe all degrees of freedom for field functions while the conventional space-time is still retained to be the background. The potential fifth variable is nothing but the proper time of particles. In response, a mass operator $(\hat{m}=-i\hbar \frac{\partial}{\partial\tau})$ should be introduced. After that, the lagrangian for free fermion fields in terms of five independent variables and mass operator is written down. By applying the gauge principle, three kinds of vector gauge couplings and one kind of scalar gauge coupling are naturally introduced. In the current scenario, the mass spectrum for all fundamental particles is accounted for in principle by solving the eigenvalue of mass operator under the function of all kinds of interactions. Moreover, there no any auxiliary mechanism including spontaneous symmetry breaking get involved in the model. Therefore, traditional problems in the standard model such as the vacuum energy problem are removed from our model, as well as the hierarchy problem on the mass spectrum for fundamental particles.\\
PACS number(s): 03.65.-w, 12.60.Cn\\
KEY WORDS: Mass Operator, Gauge Field, Scalar Coupling
\end{abstract}

\maketitle
\section{Introduction}
Recently, the ATLAS and CMS collaborations have announced the discovery of a 125Gev Higgs-like boson\cite{atlas,cms}. As we all know, Higgs particle and Higgs mechanism are the key part of the standard model. The discovery of this new boson opens a new window in exploring the deepest secret in particle physics. But the result is still preliminary. What is this new boson and what real role should it play in particle physics are still open questions.
On the other hand, in standard model Higgs mechanism predicts the existence of non-zero vacuum energy density owing to the spontaneous symmetry breaking on the vacuum of quantum field. But such a non-zero vacuum energy is not favored by theoretical physicists since it appeared. How to get the vacuum energy back to zero is a traditional puzzle. Moreover, if we apply the vacuum energy density into the dynamics of our universe, it immediately arouses a serious hierarchy problem. This is the cosmological constant problem\cite{supernova,wmap,weinberg,peebles}. Therefore, establishing a new model without Higgs mechanism, but containing a scalar coupling is an attractive research topic.

At the beginning of this paper we make a bold but rational proposal that on the premise of retaining four-dimensional conventional space-time, an additional fifth independent variable may be essential to be introduced in describing all degrees of freedom for field functions. That is the coordinate of proper time in probability distributed space. The main reason is that the production and annihilation of field particles may also be embodied in the field functions by an existence of a probability distribution on the different values of proper time coordinate. Correspondingly, we introduce a new relevant operator---mass operator in constructing the lagrangian of five-variable fermion fields. After that we perform local gauge transformations on new formalized lagrangian, then the introduction of a new gauge coupling---scalar gauge coupling is indispensable. Consequently, a five-variable gauge field theory is presented within a concise form and its few parameters may provide us with strong predictive power in further research.

\section{Five-variable Probability Distribution Functions and Mass Operator}
In the past one hundred years, the special theory of relativity and quantum
mechanics have become the most important foundations for modern physics. Both of them are well defined and
organized into logic systems. To start with, we investigate the kinematical relationship in special theory of relativity. As it is well known, the Lorentz-invariant space-time interval is given by(using natural units: $c=1$)
\begin {eqnarray}
-dS^2=-dt^2+d{\bf x}^2.
\end {eqnarray}
In parallel, the energy-momentum relation for a single particle can be written as
\begin {eqnarray}
-m^2=-E^2+{\bf p}^2.
\end {eqnarray}
On the other hand, the correspondence between kinematical quantities
and their operators in quantum mechanics is shown as
\begin {eqnarray}
E&\longrightarrow& i\hbar\frac{\partial}{\partial t}=\hat{E};\\
p_i&\longrightarrow& -i\hbar\frac{\partial}{\partial x^i}=\hat{p}_i.
\end {eqnarray}
Comparing the equation (1-2) with (3-4), we suppose that there may be an operator corresponds to the mass of particles. on the analogy of correspondences (3-4), it is natural to suppose that
\begin {eqnarray}
m\longrightarrow -i\hbar\frac{\partial}{\partial \tau}=\hat{m},
\end {eqnarray}
where the operator $\hat{m}$ is just the mass operator which is Lorentz-invariant and will be introduced into the quantum field theory.

Why should we introduce a mass operator? There are two motivations. As far as we know, the mass of fundamental particles are vastly different. There is a hierarchy problem. It is unnatural if we interpret all masses of fundamental particles by introducing just one single parameter such as one energy scale, on which the symmetry of vacuum breaks. Conversely, if there is a mass operator which is responsible to the generating of mass spectrum, it is possible to achieve a more natural picture to understand the mass spectrum for fundamental particles by resorting to the structure of couplings and interactions. This is the first motivation to introduce the mass operator.

The second motivation comes from the needs of describing full degrees of freedom in quantum field theory. As we know, since the classical mechanics is of determinism, the state of motion for any single particle is originally described by ${\bf r}(t)$=$\{x(t),y(t),z(t)\}$, which is determinable by solving the dynamical equation. Therefore, $\{x(t),y(t),z(t)\}$ can be regarded as a fundamental function for classical particles. In this sense, there are substantially three independent variables required to describe the state of motion for classical particles. However, in the scenario of quantum mechanics, the fundamental function for quantum particles is changed to be the wave function $\psi(x,y,z,t)$, which is also determinable by solving the dynamical equation in quantum mechanics, such as Schrodinger equation. Therefore, the independent variables required by the wave function in quantum mechanics are changed to be four-dimensional coordinates of the background space-time $\{t,x,y,z\}$. In other words, though a free particle has only three degrees of freedom in classical mechanics since it is of determinism, but its wave function under the framework of quantum mechanics actually has four independent variables since the wave function only describes a probability distribution of the particle. Furthermore, when the quantum mechanics is extended into quantum field theory, in the same way we should add a new independent variable into the function of field quantities, owing to the universal existence of production and annihilation of particles in quantum field theory. This kind of variable required by probability distribution functions should also be universal and qualified to describe the production and annihilation information for particles. A most natural choice is the proper time of particles, since the proper time is a universal variable in quantum field theory and not related to the specific features of particles. In some sense, the proper time can be regarded as an independent new coordinate of probability distributed space for field functions. Besides, the proper time coordinate can be used to define mass operator in the spirit of quantum mechanics. The introduction of proper time coordinate is also in a good correspondence with the mass operator for their Lorentz invariance.

Under macroscopic conditions, we know the state of motion for a classical particle is described by space-time coordinates, which satisfy the kinematical constraint relation from the special theory of relativity,
\begin {eqnarray}
d\tau^{2}=-d{\bf x}^{2}+dt^{2}.
\end {eqnarray}
Meanwhile, the invariant interval in background space-time of classical particles is given by $dS^{2}=-d{\bf x}^{2}+dt^{2}$. But at microscopic scales, the state of quantum particle is described by a distribution of probability. Moreover, the coordinates in probability distributed space are no longer constrained by above kinematical constraint relation (6). Therefore, in mathematics we are able to introduce a very simple convention on the invariant interval in five-variable probability distributed space,
\begin {eqnarray}
d\Omega=-d{\bf x}^{2}+dt^{2}-d\tau^{2}.
\end {eqnarray}
Strictly speaking, we believe that microscopic particles substantially obey the kinematical constraint relation from special theory of relativity, and the probability distribution interpretation for the wave function should be regarded as a phenomenological effective description of the state of motion for microscopic particles. Therefore, the equation (7) will not cause a new kinematical constraint relation for microscopic particles but substantially a mathematical convention for probability distributed space. In fact, we can also see from the following discussion, by introducing the convention (7), the mathematical form of the lagrangian can be simplified and symmetrized to simulate the four-dimensional covariant form in special theory of relativity. Consequently,
the vector in field function's five-variable probability distributed space can be written down as a five-component form,
\begin {eqnarray}
x_{a}=(x,y,z,t,\tau),&&a=1,2,3,4,5.
\end {eqnarray}
Due to the invariant interval in five-variable probability distributed space (7), we further introduce a metric tensor in analogy to the four-dimensional covariant formulism in special theory of relativity. The generalized metric is
\begin {eqnarray}
g^{ab}=diag(-1,-1,-1,+1,-1).
\end {eqnarray}
The five-component gradient operator is therefore introduced as
\begin {eqnarray}
\partial_{a}&=&(\frac{\partial}{\partial x},\frac{\partial}{\partial y},\frac{\partial}{\partial z},
\frac{\partial}{\partial t},\frac{\partial}{\partial \tau})=(\partial_{\mu},\partial_{\tau})      \cr
\partial_{a}\partial^{a}&=&-\frac{\partial^{2}}{\partial (x^{i})^{2}}+\frac{\partial^{2}}{\partial t^{2}}-\frac{\partial^{2}}{\partial \tau^{2}}, i=1,2,3.
\end {eqnarray}

\section{ Free Dirac Equation with Mass Operator}
After the mass operator is introduced, we investigate the commutation relation of their coefficients. The most natural requirement to these operators is that in free fermion field theory, the energy-momentum relation from special theory of relativity should be formally retained for their corresponding operators. According to the derivation of original Dirac equation, we also apply Fourier expansion method to a free fermion field function and write down a plane wave function as its general formula,
\begin {eqnarray}
\psi\propto e^{\frac{i}{\hbar}({\bf p}\cdot {\bf x}-Et+m\tau)}.
\end {eqnarray}
Then an operators' relation being analogous to Klein-Gordon equation should be obeyed if we want to construct a quantum theory in terms of operators. It is shown by
\begin {eqnarray}
(\hat{E}^{2}-c^{2}\hat{p}^{2}-c^{4}\hat{m}^{2})\psi=0.
\end {eqnarray}
But in the spirit of Dirac's equation, the relativistic equation of free fermion should be constructed by linear operators and they must be first order of derivatives. Hence a relativistic equation of free fermion field is also assumed to be
\begin {eqnarray}
[\chi(i\hbar\frac{\partial}{\partial t})-c\alpha_{j}(-i\hbar\frac{\partial}{\partial x^{j}})-c^{2}\beta(-i\hbar\frac{\partial}{\partial\tau})]\psi=0.
\end {eqnarray}
Multiplying  $[\chi(i\hbar\frac{\partial}{\partial t})+c\alpha_{j}(-i\hbar\frac{\partial}{\partial x^{j}})+c^{2}\beta(-i\hbar\frac{\partial}{\partial\tau})]$ on both sides of the equation (13), we have
\begin {eqnarray}
[\chi\hat{E}+c\alpha_{j}\hat{p}_{j}+c^{2}\beta\hat{m}][\chi\hat{E}-c\alpha_{j}\hat{p_{j}}-c^{2}\beta\hat{m}]\psi=0.
\end {eqnarray}
By making a simple expansion on the left hand side of the equation (14), we further have
\begin {eqnarray}
&[\chi^{2}\hat{E}^2-c^2\alpha_i\alpha_j\hat{p}_i\hat{p}_j+c(\alpha_i\chi\hat{p}_i\hat{E}-\chi\alpha_i\hat{E}\hat{p}_i)+\cr
&c^2(\beta\chi\hat{m}\hat{E}-\chi\beta\hat{E}\hat{m})-c^3(\alpha_i\beta\hat{p}_i\hat{m}+\beta\alpha_i\hat{m}\hat{p}_i)-\cr
&c^4\beta^{2}\hat{m}^2]\psi=0.
\end {eqnarray}
To obtain the Klein-Gordon-like relation shown as the equation (12), the commutation
relation between energy-momentum operators and mass operator is required to be
\begin {eqnarray}
\hat{m}\hat{E}-\hat{E}\hat{m}=0;\cr
\hat{m}\hat{p}_i-\hat{p}_i\hat{m}=0.
\end {eqnarray}
Above commutation relation is assigned on the basis of two reasons. Firstly, it is in analogy to the commutation relation between energy operator and momentum operator: $\hat{E}\hat{p}_i-\hat{p}_i\hat{E}=0$, which is adopted in the derivation of original Dirac equation. Secondly, the original Dirac equation has achieved a great success in both theory and experiments, while the mass term in original Dirac equation is always expressed by a real parameter. Thus in the original quantum field theory where the mass operator has not been introduced, the mass parameter is commutable with any operator. In light of this fact, we should also assume the valid of (16) from the viewpoint of phenomenology so as not to arouse an immediate conflict after we introduce a mass operator into quantum field theory. Besides, it can also be understood from the viewpoint of physics. Since the mass operator is stemmed from the inherent degree of freedom for particles, while the definition of energy operator and momentum operator are respectively based on the background time and background space, all these degrees of freedom are completely orthogonal. Therefore, their corresponding operators are commutable with each other. Based on above commutation relations between energy operator, momentum operator and mass operator, we obtain the commutation relation between their coefficients,
\begin {eqnarray}
(\alpha_1)^2=(\alpha_2)^2=(\alpha_3)^2=\beta^2=\chi^{2}=1; \cr
\alpha_i\alpha_j+\alpha_j\alpha_i=0;\cr
\alpha_i\beta+\beta\alpha_i=0; \cr
\alpha_i\chi-\chi\alpha_i=0;\cr
\chi\beta-\beta\chi=0.
\end {eqnarray}
Recall the mathematical properties of Dirac matrices, we can write down the most simple solution as,
\begin {eqnarray}
&&\alpha_i=\left(\begin{array} {cc}0&\sigma_{i} \\ \sigma_{i}& 0\end{array}\right),\gamma_i=-\beta\alpha_i=\left(\begin{array} {cc}0&-\sigma_{i} \\ \sigma_{i}& 0\end{array}\right),\cr
&&\beta=\left(\begin{array} {cc}I&0 \\ 0&-I\end{array}\right), \gamma_4=\beta,\cr
&&\chi=1.
\end {eqnarray}
Here $\sigma_{i}$ are the Pauli matrices. Correspondingly, we define a new combined representation for Dirac matrices,
\begin {eqnarray}
\gamma_a=(\gamma_{i},\gamma_{4},-I),&& a=1,2,3,4,5.
\end {eqnarray}
Now we choose the natural units: $c=\hbar=1$, so the equation (13) can be rewritten as
\begin {eqnarray}
(i\gamma^{\mu}\partial_{\mu}+i\partial_{\tau})\psi=(i\gamma^{a}\partial_{a})\psi=0
\end {eqnarray}
In this case, the lagrangian of free fermion field in the five-component representation is obtained
\begin {eqnarray}
L_{0}=\bar{\psi}\gamma^{a}\partial_{a}\psi.
\end {eqnarray}

\section{Electromagnetic Gauge Field Theory }
In analogy to the standard model, the electromagnetic gauge field can also be introduced here by considering a generalized
local $U(1)$ gauge transformation with maximized degrees of freedom in five-variable probability distributed space,
\begin {eqnarray}
\psi\longrightarrow \psi'=e^{i\alpha(x,y,z,t,\tau)}\psi,
\end {eqnarray}
here $\tau$ is the proper time of field particles and regarded as an original variable in quantum field functions.
According to the requirement of gauge invariance, we must introduce a gauge field $A_{a}$ which is assumed to satisfy the following gauge transformation,
\begin {eqnarray}
A_{a} \longrightarrow A'_{a}=A_{a}-\frac{1}{e}\partial_{a}\alpha(x,y,z,t,\tau).
\end {eqnarray}
As we have known, the gauge for electromagnetic field is given by
\begin {eqnarray}
A_{\mu} \longrightarrow A_{\mu}'=A_{\mu}-\frac{1}{e}\partial_{\mu}\alpha(x,y,z,t).
\end {eqnarray}
Therefore, when the gauge transformation is extended from four dimensions of conventional space-time into five-variable probability distributed space, not only the classical electromagnetic field is introduced, but also a new scalar field $\phi(x,y,z,t,\tau)$ get involved to describe the fifth component of $A_{a}$. In the current model, such a scalar field is naturally introduced by the gauge transformation, so it predicts a new kind of gauge coupling---scalar gauge coupling. The five-component representation for $A_{a}$ is
\begin {eqnarray}
A_{a}=(A_{1},A_{2},A_{3},\varphi,\phi)=(A_{\mu},\phi), a=1,2,3,4,5.
\end {eqnarray}
The gauge field tensor of $U(1)$ is given by
\begin {eqnarray}
F_{ab}=\partial_{a}A_{b}-\partial_{b}A_{a}.
\end {eqnarray}
Then the lagrangian of free $U(1)$ gauge field is obtained
\begin {eqnarray}
L_{0}&=&-\frac{1}{4}F_{ab}F^{ab}  \cr
&=&-\frac{1}{4}(F_{\mu\nu}F^{\mu\nu}+F_{\mu5}F^{\mu5}+F_{5\nu}F^{5\nu}+F_{55}F^{55}) \cr
&=&-\frac{1}{4}(F_{\mu\nu}F^{\mu\nu}+2F_{\mu5}F^{\mu5}).
\end {eqnarray}
In fact, the mass term of vector gauge field and the kinetic energy term of scalar gauge field have been
contained in above equation. It can be shown by making an expansion on the second term,
\begin {eqnarray}
F_{\mu5}F^{\mu5}=(\partial_{\mu}A_{5}-\partial_{5}A_{\mu})(\partial^{\mu}A^{5}-\partial^{5}A^{\mu})\cr
=(\partial_{\mu}\phi-\partial_{\tau}A_{\mu})(-\partial^{\mu}\phi-\partial^{\tau}A^{\mu})\cr
=-\partial_{\mu}\phi\partial^{\mu}\phi+2\partial_{\tau}A_{\mu}\partial^{\mu}\phi+\partial_{\tau}A_{\mu}\partial^{\tau}A^{\mu}.
\end {eqnarray}
Therefore, in the five-component representation of quantum field theory, kinetic energy term and mass term are form unified.
On the other hand, owing to $U(1)$ gauge symmetry, the kinetic energy term of fermion field can be extended into
\begin {eqnarray}
\bar{\psi}\gamma^{a}\partial_{a}\psi \rightarrow \bar{\psi}\gamma^{a}(\partial_{a}-ieA_{a})\psi.
\end {eqnarray}
We define a five-component covariant derivative for $U(1)$ gauge symmetry,
\begin {eqnarray}
D_{a}\equiv \partial_{a}-ieA_{a}.
\end {eqnarray}
Consequently, the complete lagrangian for fermion field which is invariant under a generalized $U(1)$ gauge transformation (22) is carried out in a concise expression,
\begin {eqnarray}
L=\bar{\psi}\gamma^{a}D_{a}\psi-\frac{1}{4}F_{ab}F^{ab}.
\end {eqnarray}
It is remarkable that in above expression the mass terms of fermion field and vector gauge field have both been included.

\section{ Non-Ablian Gauge Field Theory }
For simplicity, in this case we only take $SU(2)$ gauge transformation for example. Similarly, a generalized $SU(2)$ local gauge transformation with maximized degrees of freedom is given by
\begin {eqnarray}
\psi\longrightarrow \psi'=e^{\frac{i}{2}\sigma_{j}\alpha^{j}(x,y,z,t,\tau)}\psi,
\end {eqnarray}
here ${\sigma}^j$ is the $j$-th generator of $SU(2)$. The above gauge transformation has wholly contained conventional $SU(2)$
gauge transformation as it is in standard model. But at the present case, the five-component gauge field introduced by $SU(2)$ should be written as
\begin {eqnarray}
B^{i}_{a}=(B^{i}_{\mu},\phi^{i}=\phi), && a=1,2,3,4,5.
\end {eqnarray}
Here we may assume that all scalar fields introduced by different gauge symmetries correspond to the same one ($\phi^{i}=\phi$) since they are responsible to generate a unique mass spectrum. Then in the case of $SU(2)$ gauge symmetry, the five-component covariant derivative and gauge field tensor are given by
\begin {eqnarray}
D_{a}&=&\partial_{a}-\frac{i}{2}g\sigma^{i}B^{i}_{a};\cr
F^{i}_{ab}&=&\partial_{a}B^{i}_{b}-\partial_{b}B^{i}_{a}+g\epsilon^{ijk}B^{j}_{a}B^{k}_{b}.
\end {eqnarray}
The lagrangian of the free fermion field and its $SU(2)$ gauge coupling can be expanded as
\begin {eqnarray}
\bar{\psi}\gamma^{a}D_{a}\psi=\bar{\psi}\gamma^{\mu}\partial_{\mu}\psi+\bar{\psi}\partial_{\tau}\psi
-\frac{i}{2}g\bar{\psi}\gamma^{\mu}\sigma^{i}B^{i}_{\mu}\psi-\frac{i}{2}g\bar{\psi}\sigma^{i}\phi^{i}\psi.
\end {eqnarray}
It is obvious that above expression has included the mass term of the fermion field and its couplings with vectors and scalar. Analogously, we can also obtain the mass term of gauge fields and their self-interactions,
\begin {eqnarray}
&&-\frac{1}{4}F^{i}_{ab}F^{iab}\cr
&=&-\frac{1}{4}(\partial_{a}B^{i}_{b}-\partial_{b}B^{i}_{a}+g\epsilon^{ijk}B^{j}_{a}B^{k}_{b})
(\partial^{a}B^{ib}-\partial^{b}B^{ia}+g\epsilon^{ijk}B^{ja}B^{kb})\cr
&=&[-\frac{1}{2}(\partial_{\mu}B^{i}_{\nu}\partial^{\mu}B^{i\nu}-\partial_{\mu}B^{i}_{\nu}\partial^{\nu}B^{i\mu}
-\partial_{\mu}\phi^{i}\partial^{\mu}\phi^{i}+2\partial_{\mu}\phi^{i}\partial_{\tau}B^{i\mu}\cr
&&+\partial_{\tau}B^{i}_{\mu}\partial^{\tau}B^{i\mu})]+[-g\epsilon^{ijk}\partial_{\mu}B^{i}_{\nu}B^{j\mu}B^{k\nu}+g\epsilon^{ijk}\partial_{\mu}\phi^{i}B^{j\mu}\phi^{k}\cr
&&+g\epsilon^{ijk}\partial_{\tau}B^{i}_{\mu}\phi^{j}B^{k\mu}-g\epsilon^{ijk}\partial_{\tau}\phi^{i}\phi^{j}\phi^{k}]\cr
&&+[-\frac{1}{4}g^{2}\epsilon^{ijk}\epsilon^{ilm}B^{j}_{\mu}B^{k}_{\nu}B^{l\mu}B^{m\nu}
+\frac{1}{4}g^{2}\epsilon^{ijk}\epsilon^{ilm}B^{j}_{\mu}\phi^{k}B^{l\mu}\phi^{m}\cr
&&+\frac{1}{4}g^{2}\epsilon^{ijk}\epsilon^{ilm}\phi^{j}B^{k}_{\mu}\phi^{l}B^{m\mu}
-\frac{1}{4}g^{2}\epsilon^{ijk}\epsilon^{ilm}\phi^{j}\phi^{k}\phi^{l}\phi^{m}].
\end {eqnarray}
Finally, we write down the complete lagrangian for fermion field under $SU(2)$ gauge symmetry,
\begin {eqnarray}
L=\bar{\psi}\gamma^{a}D_{a}\psi-\frac{1}{4}F^{i}_{ab}F^{iab}.
\end {eqnarray}
From the expanded lagrangian for gauge fields (36) we find that the mass term of the gauge scalar field is absent. This is the biggest problem currently confronted in this preliminary model and which must be reconciled with the experiments, since an excess of events has been observed in LHC's experiment which indirectly shows that the mass of new discovered Higgs-like boson is about 125GeV.

\section{Gauge Field Theory with 3+1 Gauge Interactions}
As we know, the electromagnetic interaction, weak interaction and strong interaction actually correspond to three different gauge symmetries:
$U(1)$, $SU(2)$ and $SU(3)$ respectively. All these gauge symmetries can be written in a unified form as $U_{Y}(1)\bigotimes SU_{L}(2)\bigotimes SU_{C}(3)$. But since we introduce the mass operator, a new scalar gauge coupling should be added by the gauge principle besides above three fundamental interactions. Under the integrated gauge symmetry $U_{Y}(1)\bigotimes SU_{L}(2)\bigotimes SU_{C}(3)$, the gauge transformation of fermion field can be written down as
\begin {eqnarray}
\psi\rightarrow \psi'=e^{i[\alpha(x,y,z,t,\tau)-\frac{1}{2}\sigma^{j}\alpha^{j}(x,y,z,t,\tau)-\lambda^{I}\alpha^{I}(x,y,z,t,\tau)]}\psi.
\end {eqnarray}
According to the gauge principle, all five-component gauge fields are given by
\begin {eqnarray}
A_{a}=(A_{\mu},\phi);\cr
B^{i}_{a}=(B^{i}_{\mu},\phi^{i}=\phi), && i=1,2,3;\cr
C^{I}_{a}=(C^{I}_{\mu},\phi^{I}=\phi), && I=1,2...,8.
\end {eqnarray}
The gauge freedom for gauge fields are required to be
 \begin {eqnarray}
A_{a}\rightarrow A'_{a}&=&A_{a}-\frac{1}{g_{1}}\partial_{a}\alpha(x,y,z,t,\tau);\cr
B^{i}_{a}\rightarrow B'^{i}_{a}&=&B^{i}_{a}+\epsilon^{ijk}\alpha^{j}(x,y,z,t,\tau)B^{k}_{a}\cr
&& -\frac{1}{g_{2}}\partial_{a}\alpha^{i}(x,y,z,t,\tau);\cr
C^{I}_{a}\rightarrow C'^{I}_{a}&=&C^{I}_{a}+\Gamma^{IJK}\alpha^{J}(x,y,z,t,\tau)C^{K}_{a}\cr
&& -\frac{1}{g_{3}}\partial_{a}\alpha^{I}(x,y,z,t,\tau).
\end {eqnarray}
In a similar way, the five-component covariant derivative and gauge field tensors are defined as
\begin {eqnarray}
D_{a}&=&\partial_{a}-ig_{1}A_{a}-\frac{i}{2}g_{2}\sigma^{i}B^{i}_{a}-ig_{3}\lambda^{I}C^{I}_{a};\cr
F_{ab}&=&\partial_{a}A_{b}-\partial_{b}A_{a};\cr
F^{i}_{ab}&=&\partial_{a}B^{i}_{b}-\partial_{b}B^{i}_{a}+g_{2}\epsilon^{ijk}B^{j}_{a}B^{k}_{b};\cr
F^{I}_{ab}&=&\partial_{a}C^{I}_{b}-\partial_{b}C^{I}_{a}+g_{3}\Gamma^{IJK}C^{J}_{a}C^{K}_{b}.
\end {eqnarray}
Finally, under the integrated gauge symmetry, the complete lagrangian of gauge field theory is carried out as
\begin {eqnarray}
L=\bar{\psi}\gamma^{a}D_{a}\psi-\frac{1}{4}F_{ab}F^{ab}-\frac{1}{4}F^{i}_{ab}F^{iab}-\frac{1}{4}F^{I}_{ab}F^{Iab}.
\end {eqnarray}
So far there are three kinds of vector gauge couplings and one kind of scalar gauge coupling presented in our model by the principle of lagrangian invariant under local gauge transformation. Among them, three kinds of vector couplings can still correspond to the well known interactions. But as for the new scalar field we find it is in coupling with almost all fundamental particles. Therefore, such a scalar gauge coupling is virtually a universal coupling. We suppose this kind of scalar coupling may be related to the gravitational interaction although it is introduced by the gauge symmetry. In addition, there also exists another possibility. The scalar field introduced by mass operator is originated from the requirement that the proper time of particles should be reflected in the distribution of probability. In other words, the existence of this kind of scalar coupling may result in the change on the value of particles' proper time. Therefore, this scalar field may also depict a kind of interaction which arouses the internal evolution or spontaneous decay of particles. Anyhow, it deserves further investigations.

\section{Conclusion}
In the standard model of particle physics, we know that the mass of fundamental particles is provided by a spontaneous symmetry breaking on complex scalar doublet Higgs field and this approach is the so-called Higgs mechanism. But there are several questions existing for such a mechanism. First, Higgs complex scalar and its Yukawa coupling with other fundamental particles are put by hand into the lagrangian. This is unnatural. Second, the symmetry breaking in Higgs mechanism will cause a large vacuum energy density and we now know that it is far larger than the upper limit imposed by the current cosmological observations. On the other hand, in Higgs mechanism, the Higgs scalar is required to couple with almost all fundamental particles. In fact such a universal coupling should only be possessed by the gravitational interaction. Besides, both mass and energy are naturally related to the gravity. Therefore, we have ever introduced a gravitational Higgs mechanism after we generalized Brans-Dicke theory \cite{scalar gravity}. In that model, according to the change of cosmological curvature we obtain a running vacuum energy density and a running mass spectrum for fundamental particles. The speed of running is survivable under the constraints from particle physics experiments, due to a slow expanding speed of our universe at the present time.

Though gravitational Higgs mechanism is very attractive for its sound physical motivations, it still has a serious hierarchy problem as it does in standard model. The main difficulty is that the mass spectrum of all fundamental particles is generated by a single parameter in any improved models as long as the mechanism of spontaneous symmetry breaking is retained. On the other hand, how about a non-gravitational mass generating mechanism? And then what is its natural physical picture? In this paper, we suppose that the essence of the mass is the bound state of energy. So the most natural way to understand the mass spectrum for fundamental particles is to solve the eigen equation of mass operator. To define a mass operator, we make a bold but rational assumption in this paper. We extend the number of independent variables of field functions from four to five in considering that the production and annihilation of field particles exist universally. So the proper time of the field particle is treated as an independent coordinate in probability distributed space. Just like other coordinates $\{x,y,z,t\}$ of conventional space-time, the field functions should describe a probability distribution also on the proper time dimension. The independence for the proper time coordinate is actually traceable. In classical mechanics, the spatial coordinates $x,y,z$ for any particle at any time must be uniquely determined by Newton's dynamical law which is of determinism. But in the wave function description of quantum mechanics, $\{x,y,z,t\}$ appear as independent variables since the wave function describes a probability distribution on four degrees of freedom. By that analogy, the coordinate of proper time should also be introduced as an independent variable into field function's description and the field function totally describes a probability distribution on five degrees of freedom. In a word, though the space-time background is kept to be four-dimensional, the introducing of five independent variables into field functions is not only permitted by the special language of quantum physics describing probability, but may also be required by the production and annihilation phenomena in quantum field theory. On this basis, we further introduce a mass operator by making an analogy between the kinematical variables in special theory of relativity and the operators in quantum mechanics. After that, by using the gauge principle on the five-variable free fermion field theory, we can find not only the electromagnetic interaction, weak interaction and strong interaction are introduced, but also a new universal scalar gauge coupling is introduced.

In our model, almost all fundamental particles' mass terms can be naturally included into the theory by the form of mass operator. Therefore, in principle the mass of fundamental particles is determined by solving the eigen equation of the mass operator under the existence of all couplings. To solve the eigen equation of mass operator, an integrated consideration of the whole gauge field theory is indispensable so there is some difficult from mathematics in solving this problem. But there is no any auxiliary field put into the theory by hand and also no vacuum energy density exists. It is hopeful to solve the traditional cosmological constant problem. Besides, since the mass spectrum for fundamental particles is now determined by the mass operator, it is also meaningful to naturally explain the hierarchy problem.
Finally, although in this paper the discussion is not focused on the constructing of an exact model since the neutrino physics is still an open question, we still present a novel mechanism by which the mass operator and scalar gauge coupling can be naturally introduced.

\section{Acknowledgments:} I would like to thank all my friends who have provided their
constructive criticisms and comments. This work has been supported from the Nature Science Foundation of ZheJiang Province under the grant number Y6110778. This research was supported in part by the Project of Knowledge Innovation Program (PKIP) of Chinese Academy of Sciences, Grant No. KJCX2.YW.W10. This article is dedicated to my sister Ms. Chen ShuXia.

\end{document}